\documentclass[12pt]{article}
\usepackage{amssymb,amsmath,epsfig}

\begin{document}

\title{\bf Energy of Bardeen Model Using Approximate Symmetry Method}

\author{M. Sharif \thanks{msharif@math.pu.edu.pk} and Saira Waheed
\thanks{sairawaheed\_50@yahoo.com}\\
Department of Mathematics, University of the Punjab,\\
Quaid-e-Azam Campus, Lahore-54590, Pakistan.}

\date{}

\maketitle
\begin{abstract}
In this paper, we investigate the energy problem in general
relativity using approximate Lie symmetry methods for differential
equations. This procedure is applied to Bardeen model (the regular
black hole solution). Here we are forced to evaluate the third-order
approximate symmetries of the orbital and geodesic equations. It is
shown that energy must be re-scaled by some factor in the
third-order approximation. We discuss the insights of this
re-scaling factor.
\end{abstract}
{\bf MSCS:} 83C40; 70S10

\section{Introduction}

Energy-momentum is an important conserved quantity in any physical
theory whose definition has been under investigation for a long time
from the general relativity viewpoint. Unfortunately, there is still
no generally accepted definition of energy-momentum for
gravitational fields. The main problem is with the expression
defining the gravitational field energy part. To obtain a meaningful
expression for energy, momentum and angular momentum for a general
relativistic system, Einstein proposed an expression, called
Einstein's energy-momentum complex \cite{2}. After that many
complexes have been found, for instance, Landau-Lifshitz \cite{3},
Tolman \cite{4}, Papapetrou \cite{5}, M$\o$ller \cite{6}, Weinberg
\cite{8} and Bergmann \cite{9}. Some of these definitions are
coordinate dependent while others are not. Also, some of these
expressions cannot be used to define angular momentum.

Virbhadra \cite{J15} gave marvellous idea about coincidence of some
complexes which attracted many researchers. Aguireregabiria et al.
\cite{J17} proved that Einstein, Landau-Lifshitz, Papapetrou,
Weinberg prescriptions yielded the same distribution of energy for
any metric of the Kerr-Schild class if calculations are performed in
Kerr-Schild coordinates. Vagenas \cite{J22} and Sharif \cite{J23}
proved that these four prescriptions coincide for the
$(2+1)$-dimensional rotating BTZ black hole (named after Banados,
Teitelboim and Zanelli) and regular black hole solution
respectively. Xulu \cite{J24} extended this investigation and found
the same energy distribution for the dyadosphere of the
Reissner-Nordstr$\ddot{o}$m (RN) black hole. However, all these
prescriptions have their own drawbacks.

The concept of approximate symmetry theories comes from the
combination of Lie group theory and perturbations. Using this idea
two different so-called ``approximate symmetry'' theories have been
constructed. The first theory was proposed by Baikov, Gazizov and
Ibragimov \cite{20} and the second was presented by Fushchich and
Shtelen \cite{21}. It may be mentioned here that a manifold having
no exact symmetry can possess approximate symmetries. As compared to
exact symmetry, one can acquire more interesting information from a
slightly broken (approximate) symmetry. This idea of symmetries
would be more accurate as it does not depend on pseudo-tensors and
hence agrees with the principles of general relativity. In order to
discuss the approach of approximate symmetries, we use the concept
of Lie symmetries which is substantiated in spacetime isometries as
well as in Noether symmetries.

Kara et al. \cite{22} used approximate Lie symmetry method \cite{20}
to discuss the conservation laws of the Schwarzschild spacetime.
Later, using the same procedure, in the second-order approximation,
the lost conservation laws for RN \cite{23}, Kerr-Newmann \cite{24}
and Kerr-Newmann AdS \cite{25} spacetimes were recovered and some
energy re-scaling factors were obtained. In a recent paper
\cite{25a}, we have discussed re-scaling of energy in the stringy
charged black hole solutions using approximate symmetries. In going
from the Minkowski spacetime to non-flat regular black hole
solution, the recovered conservation laws (recovered in first-order
approximation) are lost. It is expected that by minimizing the
charge, one should recover all lost conservation laws. In this
paper, we use the same procedure to formulate energy re-scaling
factor for regular black hole solution.

The format of the paper is as follows. In the next section, we
review the basic material about approximate symmetry methods for the
solution of differential equations (DEs) and also give exact
symmetries of the Minkowski spacetime and the first-order
approximate symmetries of the Schwarzschild spacetime. In section
\textbf{3}, we study approximate symmetries of the regular black
hole solution. Finally, we summarize and discuss the results.

\section{Mathematical Formalism}

Symmetry is a point transformation (a transformation which maps one
point $(x,y)$ into another $(x^{*},y^{*})$) under which the form of
DE does not change. These transformations form a group known as
point transformation group. Symmetries are important for their
direct connection with the conservation laws through the Noether
theorem \cite{27}. It has been mentioned that if, for a given system
of DEs, there is a variational principle, then a continuous symmetry
invariant under the action of the given functional provides
corresponding conservation law \cite{28,30}.

A system of $p$ ODEs, each of $n^{th}$-order \cite{17,31}
\begin{equation}\label{1}
\textbf{E}_{\alpha}(s;\textbf{x}(s),\textbf{x}'(s),\textbf{x}''(s),...,\textbf{x}^{(n)}(s))=0,\quad
(\alpha=1,2,3..,p)
\end{equation}
(here $\textbf{x}$ is dependent variable, $s$ is independent
variable and $\textbf{x}',~ \textbf{x}'',..., \textbf{x}^{(n)}$
denote the first, second and so on $n^{th}$-order derivative of
$\textbf{x}$ w.r.t. $s$) under the point transformation:
$(s,\textbf{x})\longrightarrow (\xi(s,\textbf{x}),\eta
(s,\textbf{x}))$ can admit a symmetry generator
\begin{equation}\label{2}
\textbf{X}=\xi(s,\textbf{x})\frac{\partial}{\partial
s}+\eta^{\alpha}(s,\textbf{x})\frac{\partial}{\partial
\textbf{x}^{\alpha}}
\end{equation}
if and only if, on the solution of the ODEs,
$\textbf{E}_{\alpha}=0$, the following symmetry condition
\begin{equation}\label{3}
\textbf{X}^{[n]}(\textbf{E}_{\alpha})\mid_{\textbf{E}_{\alpha=0}}=0
\end{equation}
is satisfied. Here $n^{th}$-order extension of the symmetry
generator given by Eq.(\ref{2}) can be written as
\begin{eqnarray}\label{4}
\textbf{X}^{[n]}&=&\xi(s,\textbf{x})\frac{\partial}{\partial
s}+\eta^{\alpha}(s,\textbf{x})\frac{\partial}{\partial
\textbf{x}}+\eta^{\alpha}_{,s}(s,\textbf{x},\textbf{x}')\frac{\partial}{\partial
\textbf{x}^{\alpha'}
}+...\nonumber\\
&+&\eta^{\alpha}_{,(n)}(s,\textbf{x},\textbf{x}',...,\textbf{x}^{(n)})\frac{\partial}{\partial
\textbf{x}^{\alpha(n)}}
\end{eqnarray}
and the corresponding prolongation coefficients are
\begin{equation}\label{5}
\eta^{\alpha}_{,s}=\frac{d\eta^{\alpha}}{ds}-\textbf{x}^{\alpha'}\frac{d\xi}{ds},\quad
\eta^{\alpha}_{,(n)}=\frac{d\eta^{\alpha}_{,(n-1)}}{ds}-
\textbf{x}^{\alpha(n)}\frac{d\xi}{ds};\quad n\geq2.
\end{equation}
The $k^{th}$-order approximate symmetries of a perturbed ODE
\begin{equation}\label{6}
\textbf{E}=\textbf{E}_{0}+{\epsilon}\textbf{E}_{1}+\epsilon^{2}\textbf{E}_{2}+...
+\epsilon^{k}\textbf{E}_{k} +O(\epsilon^{k+1});\quad \epsilon \in
R^{+}
\end{equation}
is given by the generator
\begin{equation}\label{7}
\textbf{X}=\textbf{X}_{0}+{\epsilon}\textbf{X}_{1}+\epsilon^{2}\textbf{X}_{2}+...
+\epsilon^{k}\textbf{X}_{k}
\end{equation}
if the following symmetry condition holds \cite{32}
\begin{eqnarray}\label{8}
\textbf{X}E&=&[(\textbf{X}=\textbf{X}_{0}+{\epsilon}\textbf{X}_{1}
+\epsilon^{2}\textbf{X}_{2}+... +\epsilon^{k}\textbf{X}_{k})
(\textbf{E}=\textbf{E}_{0}+{\epsilon}\textbf{E}_{1}\nonumber\\
&+&\epsilon^{2}\textbf{E}_{2}+...+\epsilon^{k}\textbf{E}_{k})]
\mid_{{\textbf{E}=\textbf{E}_{0}+{\epsilon}\textbf{E}_{1}
+\epsilon^{2}\textbf{E}_{2}+...+\epsilon^{k}\textbf{E}_{k}}} =
O({\epsilon}^{k+1}).
\end{eqnarray}
Here $\textbf{E}_{0}$  is the exact part of ODE,
$\textbf{E}_{1},~\textbf{E}_{2}$ are respectively the first and
second order approximate parts of the perturbed ODE and so on.
$\textbf{X}_{0}$ represents the exact symmetry generator and
$\textbf{X}_{1},~\textbf{X}_{2}$ denote respectively the first and
second order approximate parts of the symmetry generator and so on.
For the $k^{th}$-order approximate symmetry generator, the terms
involving $\epsilon^{(k+1)}$ and its higher powers can be
substituted to zero so that RHS of Eq.(\ref{8}) becomes zero. As
compared to exact symmetries, these approximate symmetries do not
necessarily form a Lie algebra rather do form the so-called
``approximate Lie algebra" (up to a specified degree of precision)
\cite{33}. The perturbed equation always admits an approximate
symmetry $\epsilon \textbf{X}_{0}$ called a trivial symmetry. If a
symmetry generator $\textbf{X}=\textbf{X}_{0}+\epsilon
\textbf{X}_{1}$ exists with $\textbf{X}_{0}\neq 0$ and
$\textbf{X}_{1}\neq k\textbf{X}_{0}$, $k$ is an arbitrary constant,
then it is called non-trivial symmetry \cite{34}.

It has been mentioned \cite{35} that $10$ Killing vectors (Appendix
\textbf{A}) of Minkowski spacetime provide conservation laws for
energy and linear momentum as well as for angular and spin angular
momentum. The algebra calculated from the geodesic equations
contains some additional symmetries which provide no conservation
law \cite{18}. The Schwarzschild spacetime has four isometries
$\textbf{Y}_{0},~\textbf{Y}_{1},~\textbf{Y}_{2},~\textbf{Y}_{3}$
which correspond to conservation laws of energy and angular momentum
only. The symmetry algebra of the Schwarzschild spacetime
(calculated through the geodesic equations) consist of four
isometries and the dilation algebra. In the limit of small mass of
point gravitating source, $\epsilon=2M$ \cite{22}, all lost
conservation laws would be recovered as first order trivial
approximate symmetries. For the orbital equation, the first-order
approximate symmetries are given in Appendix \textbf{B}.

\section{Bardeen Model}

Reissner-Nordstr$\ddot{o}$m solution is a static, spherically
symmetric and asymptotically flat solution of the Einstein-Maxwell
field equations having a singularity structure. There arise some
problems when one applies the laws of physics at the singularity
point (origin of the radial coordinate) of the RN solution. To
resolve this problem, Bardeen (1968) constructed a singularity free
solution \cite{36,37} using the energy-momentum tensor of non-linear
electrodynamics as the source of the field equations. The Bardeen
model is known as a regular black hole solution because for a
particular ratio of mass to charge, this represents a black hole and
a singularity free structure. After that many regular black hole
solutions have been discussed by various authors, for example,
\cite{38}. These metrics, being analogues to the RN metric with
magnetic charge, have a quite similar global but regular structure
as that for the RN spacetime.

\subsection{Isometry Algebra and Orbital Equation of Motion}

The line element representing Bardeen model is given by \cite{37}
\begin{equation}\label{40}
ds^{2}=[1-\frac{2mr^{2}}{(r^{2}+e^{2})^{3/2}}]dt^{2}
-[1-\frac{2mr^{2}}{(r^{2}+e^{2})^{3/2}}]^{-1}dr^{2}-r^{2}(d{\theta}^{2}+
\sin^{2}{\theta}d{\phi}^{2}),
\end{equation}
The asymptotic behavior of the solution can be observed by the
simple expansion of the term $g_{tt}$ as follows
\begin{equation*}
g_{tt}=1-\frac{2m}{r}+\frac{3me^{2}}{r^{3}}+O(\frac{1}{r^{5}})
\end{equation*}
where $m$ is the mass of the configuration. When $r\longrightarrow
\infty$, the parameter $e$ does not vanish as the Columbian term
$\frac{1}{r^{2}}$ but vanishes as the term $\frac{1}{r^{3}}$.
Therefore, one cannot recognize the parameter $e$ as the Coulomb
charge $Q$ of the RN spacetime for which $g_{tt}$ is given by
\begin{equation*}
g_{tt}=1-\frac{2m}{r}+\frac{Q^{2}}{r^{2}}.
\end{equation*}
Indeed, in the Bardeen model, the parameter $e$ represents the
magnetic charge of the non-linear self-gravitating monopole
\cite{38a} while we have taken the gravitational units, i.e.,
$G=c=1$. For $e=0$, it turns out to be the Schwarzschild spacetime
while for $m=0=e$, it becomes Minkowski spacetime. The simultaneous
solution of the Killing equations for the Bardeen model is given by
\begin{eqnarray*}
k^{0}=c_{0},~k^{1}=0,~k^{2}=c_{1}\sin\phi-c_{2}\cos\phi,~
k^{3}=\cot\theta(c_{1}\cos\phi+c_{2}\sin\phi)+c_{3}.
\end{eqnarray*}
This solution corresponds to the generators
$\textbf{Y}_{0},~\textbf{Y}_{1},~ \textbf{Y}_{2}$ and
$\textbf{Y}_{3}$ which form an algebra $so(3){\oplus}R$ implying the
conservation laws for energy and angular momentum only while the
linear and spin angular momentum conservation laws are lost. We
would check whether one can recover these lost conservation laws in
the limit of small gravitational mass and charge or not.

The set of geodesic equations for the spacetime (\ref{40}) can be
written as
\begin{eqnarray}\label{41}
&&\ddot{t}+2[\frac{m(r^{2}+e^{2})^{1/2}}{[1-\frac{2mr^{2}}
{(r^{2}+e^{2})^{3/2}}]}][\frac{(r^{3}-2re^{2})}{(r^{2}+
e^{2})^{3}}]\dot{t}\dot{r}=0,\\\nonumber
&&\ddot{r}+[\frac{mr(r^{2}-2e^{2})}{(r^{2}+e^{2})^{5/2}}]
[1-\frac{2mr^{2}}{(r^{2}+e^{2})^{3/2}}]\dot{t}^{2}\\\label{42}
&&-\frac{mr(r^{2}-2e^{2})}{(r^{2}+e^{2})^{5/2}}[1-\frac{2mr^{2}}
{(r^{2}+e^{2})^{3/2}}]^{-1}\dot{r}^{2}\nonumber\\
&&-r[1-\frac{2mr^{2}}{(r^{2}+e^{2})^{3/2}}](\dot{\theta}^{2}
+\sin^{2}\theta\dot{\phi}^{2})=0, \\\label{43}
&&\ddot{{\theta}}+\frac{2}{r}\dot{r}\dot{\theta}- \sin\theta
\cos\theta \dot{\phi}^{2}=0,\\\label{44}
&&\ddot{{\phi}}+\frac{2}{r}\dot{r}\dot{\phi}+2\cot{\theta}
\dot{\theta}\dot{\phi}=0.
\end{eqnarray}
Using the spacetime, four geodesic equations and the symmetries of
the equatorial plane, we obtain the following relativistic equation
of motion (orbital equation of motion)
\begin{equation}\label{45}
\frac{d^{2}u}{{d\phi}^{2}}+u=\frac{2mu^{2}}{(1+u^{2}e^{2})^{3/2}}+
\frac{m(1-2e^{2}u^{2})}{h^{2}(1+e^{2}u^{2})^{5/2}}+\frac{mu^{2}
(1-2e^{2}u^{2})}{(1+e^{2}u^{2})^{5/2}}
\end{equation}
where $h$ is the classical angular momentum per unit mass and
$u=\frac{1}{r}$

\subsection{Perturbation Parameters and Approximate Symmetries of
the Orbital and Geodesic Equations}

The Bardeen model is a singularity free solution and represents
black hole for the following inequality
\begin{equation}\label{46}
27e^{2}\leq 16m^{2}.
\end{equation}
Since the vanishing of mass and charge provide the Minkowski
spacetime, so the perturbation parameters are to be defined in terms
of mass and charge of the black hole to recover the lost
conservation laws. The required perturbation parameters are
\begin{equation*}
\epsilon=2m,\quad e^{2}=k\epsilon^{2};\quad 0<k\leq\frac{4}{27},
\end{equation*}
(where we have used the inequality defined in Eq.(\ref{46})).
Introducing these parameters in the Bardeen model, we obtain the
following third-order perturbed spacetime
\begin{equation*}
ds^{2}=[1-\frac{\epsilon}{r}+\frac{3k\epsilon^{3}}{2r^{3}}]dt^{2}-[1+\frac{\epsilon}{r}
+\frac{\epsilon^2}{r^2}-\frac{3k\epsilon^{3}}{2r^{3}}]dr^2-r^{2}(d{\theta}^{2}+
\sin^{2}{\theta}d{\phi}^{2}).
\end{equation*}
Retaining only the terms of first and second order in the above
equation and neglecting $O(\epsilon^{3})$, it reduces to the
second-order perturbed spacetime for the Schwarzschild solution.
Also, it reduces to the Minkowski spacetime in the limit of
$\epsilon=0$. The third-order perturbed orbital equation of motion
can be written as
\begin{equation}\label{47}
\frac{d^{2}u}{d{\phi}^{2}}+u=\epsilon(\frac{1}{2h^{2}}+\frac{3}{2}
u^{2})-\epsilon^{3}(\frac{15ku^{4}}{4}+\frac{9ku^{2}}{4h^{2}}).
\end{equation}
For this equation of motion, the exact and first-order parts of the
symmetry generator are the same as for the Schwarzschild spacetime
given by Eqs.(\ref{61})-(\ref{67}). Since there is no term quadratic
in $\epsilon$, therefore its second-order approximate symmetries are
zero. Now we calculate its third-order order approximate symmetries.
For this purpose, we apply the operator
\begin{equation*}
\textbf{X}^{[2]}=\xi(\phi,u)\frac{\partial}{\partial\phi}+\eta(\phi,u)\frac{\partial}{\partial
u}+\eta_{,\phi}(\phi,u,u')\frac{\partial}{\partial u'}
+\eta_{,\phi\phi}(\phi,u,u',u'')\frac{\partial}{\partial u''},
\end{equation*}
to Eq.(\ref{47}), then substitute the values of the prolongation
coefficients and retain only the terms involving $\epsilon^{3}$, we
obtain a system of four DEs. The simultaneous solution of these
equations is exactly the same as given by Eq.(\ref{61})-(\ref{65}).
Thus we can conclude that there exist no non-trivial third-order
approximate symmetries.

In order to recover full set of lost conservation laws, we apply the
approximate symmetry procedure to the system of geodesic equations
given by Eqs.(\ref{41})-(\ref{44}). The third-order perturbed
geodesic equations are
\begin{eqnarray}\label{48}
&&\ddot{t}+\frac{\epsilon}{r^{2}}\dot{t}\dot{r}+\frac{\epsilon^{2}}{r^{3}}\dot{t}\dot{r}
+\frac{\epsilon^{3}}{2r^{4}}(2-9k)\dot{t}\dot{r}=0,\\\nonumber
&&\ddot{r}-r(\dot{\theta}^{2}+\sin^{2}\theta\dot{\phi}^{2})
+\epsilon[\frac{1}{2r^{2}}(\dot{t}^{2}
+\dot{r}^{2})+\dot{\theta}^{2}\\\nonumber
&&+\sin^{2}\theta\dot{\phi}^{2}]-\frac{\epsilon^{2}}{2r^{3}}[\dot{t}^{2}-\dot{r}^{2}]-
\frac{\epsilon^{3}}{2r^{4}}[\frac{9k}{2}\dot{t}^{2}-(1-\frac{9k}{2})\dot{r}^{2}\\\label{49}
&&-3kr^{2}(\dot{\theta}^{2} +\sin^{2}\theta\dot{\phi}^{2})]=0,
\\\label{50} &&\ddot{{\theta}}+\frac{2}{r}\dot{r}\dot{\theta}-
\sin\theta \cos\theta \dot{\phi}^{2}=0,\\\label{51}
&&\ddot{{\phi}}+\frac{2}{r}\dot{r}\dot{\phi}+2\cot{\theta}
\dot{\theta}\dot{\phi}=0.
\end{eqnarray}
For $\epsilon=0$, this system of equations reduces to the system for
the Minkowski spacetime. For $\epsilon^{2}=0$ with $\epsilon\neq 0$,
this system results in the first-order perturbed system of geodesic
equations for the Schwarzschild spacetime. Therefore, its exact and
first-order approximate parts of the symmetry generator are the same
as for the Minkowski and Schwarzschild spacetimes respectively.

For the second-order approximate part of the symmetry generator, we
apply the second-order prolongation of the symmetry generator
$\textbf{X}^{[2]}$. This is given by
\begin{eqnarray}\nonumber
\textbf{X}^{[2]}&=& \xi\frac{\partial}{\partial
s}+\eta^{0}\frac{\partial}{\partial
t}+\eta^{1}\frac{\partial}{\partial
r}+\eta^{2}\frac{\partial}{\partial
\theta}+\eta^{3}\frac{\partial}{\partial
\phi}+\eta^{0}_{,s}\frac{\partial}{\partial
\dot{t}}+\eta^{1}_{,s}\frac{\partial}{\partial
\dot{r}}\\\label{52}&+&\eta^{2}_{,s}\frac{\partial}{\partial
\dot{\theta}}+\eta^{3}_{,s}\frac{\partial}{\partial \dot{\phi}}
+\eta^{0}_{,ss}\frac{\partial}{\partial
\ddot{t}}+\eta^{1}_{,ss}\frac{\partial}{\partial
\ddot{r}}+\eta^{2}_{,ss}\frac{\partial}{\partial
\ddot{\theta}}+\eta^{3}_{,ss}\frac{\partial}{\partial \ddot{\phi}},
\end{eqnarray}
where $\xi=\xi_{0}+{\epsilon}\xi_{1}+\epsilon^{2}\xi_{2}$ and
$\eta^{\nu}=\eta_{0}^{\nu}+{\epsilon}\eta_{1}^{\nu}+\epsilon^{2}\eta_{2}^{\nu},~
(\nu=0,1,2,3)$. The application of this operator to the second-order
perturbed system of geodesic equations (i.e.,
$\epsilon^{3}=0,~\epsilon\neq0,~\epsilon^2\neq0$) yields the system
of $60$ DEs. The solution (where we have retained only the terms
involving $\epsilon^2$) to this system of DEs is 10 Killing vectors
and the dilation algebra $d_{2}$.

For the third order approximate symmetries, we apply the above
operator $X^{[2]}$ to third-order perturbed geodesic equation given
by Eqs.(\ref{48})-(\ref{51}), where
$\xi=\xi_{0}+{\epsilon}\xi_{1}+\epsilon^{2}\xi_{2}+\epsilon^{3}\xi_{3}$
and
$\eta^{\nu}=\eta_{0}^{\nu}+{\epsilon}\eta_{1}^{\nu}+\epsilon^{2}\eta_{2}^{\nu}
+\epsilon^{3}\eta_{3}^{\nu}$. The application of the operator again
yields a system of $60$ determining equations (where we have
retained only the terms involving $\epsilon^{3}$). In the
construction of this system of DEs, we have used $10$ Killing
vectors in which $4$ symmetries (given by Eqs.(\ref{14})-(\ref{16}))
are taken as exact and the remaining $6$ (given by
Eqs.(\ref{17}))-(\ref{22})) as first-order and second-order parts of
the symmetry generators.

We have seen that out of four constants corresponding to exact
symmetry generators, two have been appeared and canceled out while
the six constants corresponding to first and second-order parts of
the symmetry generators appear. All these six constants turn out to
be zero when we solve this system of equations simultaneously and it
results in the system of DEs for the Minkowski spacetime. Its
solution provides $10$ Killing vectors and the generators
corresponding to dilation algebra $d_{2}$. This leads to no
non-trivial symmetry in third-order approximation. For this model,
the constants corresponding to dilation algebra
$\xi_{0}(s)=c_{0}s+c_{1}$ do not cancel automatically but collect
some expression for cancelation. This expression is given as follows
\begin{equation}\label{00}
1-\frac{9k}{2}=1-\frac{9e^{2}}{8m^{2}},
\end{equation}
where we have used the value of $k$. Since $\xi(s)$ is the
coefficient of $\frac{\partial}{\partial s}$ (where $s$ corresponds
to proper time) in the symmetry generator given in Eq.(\ref{2}), so
we call this factor as energy re-scaling factor of a test particle
due to the presence of charge and mass.

\section{Summary and Discussion}

This paper is devoted to study energy of the Bardeen model (regular
black hole solution) by using the approximate symmetries. These
solutions have the isometry algebra $so(3){\oplus}R$ while the
system of geodesic equations have algebra
$so(3){\oplus}R{\oplus}d_{2}$. The strength of the approximate
symmetries (i.e., perturbation parameters) provide the information
about the amount of energy contained in the spacetime field. The
second-order approximation to symmetry generator yields only the
lost conservation laws and hence no information regarding the energy
re-scaling factor is obtained. For the energy re-scaling factor, we
have to evaluate the third-order approximate symmetries.

Firstly, we have evaluated the third-order approximate symmetries of
the orbital equation of motion. The exact and the first-order parts
of the symmetry generators are the same as for the Schwarzschild
spacetime. However, the second-order approximate part of the
symmetry generator is zero. In the third-order approximation, one
can obtain the trivial symmetries only (i.e., of Minkowski
spacetime) given by Eqs.(\ref{61})-(\ref{65}).

Secondly, we have calculated the third-order approximate symmetries
of the perturbed geodesic equations. Here the exact, first-order and
second-order approximate parts of the symmetry generator turn out to
be the same as for the Schwarzschild spacetime. The third-order
approximate symmetries of the third-order perturbed geodesic
equations provide no non-trivial symmetry generator. However, we
have obtained energy re-scaling factor given by Eq.(\ref{00}). The
re-scaling factor is of great interest as it shows the reduction in
energy by the ratio of magnetic self-energy of the source to its
gravitational self-energy. For $e=0$, it vanishes and coincides with
the Schwarzschild spacetime.

When a particle is placed in the field produced by charge
gravitating source, some force is exerted on it which should be
position dependent. However it is not clear whether the modification
in mass (energy) in the presence of charge should be position
dependent. The expression for re-scaling of force calculated through
the pseudo-Newtonian formalism \cite{40} is given by
\begin{equation}\nonumber
\frac{m}{r^{2}(r^{2}+e^{2})^{5/2}}{(r^{3}-2re^{2})}
\end{equation}
It is position dependent and provides the reduction in force by a
factor depending on the ratio of the magnetic potential energy at a
distance $r$ to the rest energy of the gravitational source.
However, it is not a convenient expression for relating the magnetic
self-energy of the configuration to its gravitational self-energy.
It should be position independent where the radial parameter $r$
would be canceled out. Our obtained re-scaling factor is clearly
position independent and corresponds to change in mass (energy) due
to the presence of magnetic charge. Therefore, it provides a
physically more significant expression for relating these
self-energies as compared to the results found through different
approaches \cite{41}.

We would like to mention here that Bardeen model with magnetic
charge is the analogue of the RN spacetime with electric charge for
which, in the second-order approximation an energy re-scaling factor
\begin{equation}
1-2k=1-\frac{Q^{2}}{2M^{2}}
\end{equation}
is obtained. However, in contrast to this, we are forced to
calculate its third-order approximate symmetries because the
second-order approximation provide no information about the energy
re-scaling factor. Clearly both these scaling factors are position
independent providing more significant results. For the orbital
equation of motion, the second-order approximate symmetries are zero
(while for the RN spacetime, there exists non-zero and trivial
second-order symmetries). In literature \cite{22,23}, a difference
between the conservation laws for the full system of geodesic
equations and single orbital equation of motion is noted. This
difference also holds for the Bardeen solution. It would be
worthwhile to investigate the approximate symmetry generators and
the concept of energy re-scaling for the rotating stringy black hole
solutions which contain electric, magnetic and both charges.

\renewcommand{\theequation}{A\arabic{equation}}
\setcounter{equation}{0}
\section*{Appendix A}

The symmetry generators of the Minkowski spacetime are given as
follows:
\begin{eqnarray}\label{14}
\textbf{Y}_{0}&=&\frac{\partial}{\partial t},\quad
\textbf{Y}_{1}=\cos\phi \frac{\partial}{\partial \theta}-\cot\theta
\sin\phi\frac{\partial}{\partial \phi},\\\label{15}
\textbf{Y}_{2}&=&\sin\phi\frac{\partial}{\partial \theta}+\cot\theta
\cos\phi\frac{\partial}{\partial \phi}, \\\label{16}
\textbf{Y}_{3}&=&\frac{\partial}{\partial \phi}, \\\label{17}
\textbf{Y}_{4}&=&\sin\theta\cos\phi \frac{\partial}{\partial r}+
\frac{\cos\theta \cos\phi}{r} \frac{\partial}{\partial
\theta}-\frac{\csc\theta\sin\phi}{r}\frac{\partial}{\partial\phi},\\\label{18}
\textbf{Y}_{5}&=&\sin\theta\sin\phi \frac{\partial}{\partial
r}+\frac{\cos\theta \sin\phi}{r}\frac{\partial}{\partial \theta
}+\frac{\csc\theta
\cos\phi}{r}\frac{\partial}{\partial\phi},
\end{eqnarray}
\begin{eqnarray}\label{19}
\textbf{Y}_{6}&=&\cos\theta\frac{\partial}{\partial r}-\frac{\sin
\theta}{r}\frac{\partial}{\partial\theta}, \\\nonumber
\textbf{Y}_{7}&=&r \sin\theta \cos\phi\frac{\partial}{\partial
t}+t(\sin\theta \cos\phi\frac{\partial}{\partial r}+\frac{\cos\theta
\cos\phi}{r}\frac{\partial}{\partial\theta}
-\frac{\csc\theta\sin\phi}{r}\frac{\partial}{\partial\phi}),\\\label{20}
\\\nonumber \textbf{Y}_{8}&=&r\sin\theta
\sin\phi\frac{\partial}{\partial t}+t(\sin\theta
\sin\phi\frac{\partial}{\partial r}+\frac{\cos\theta
\sin\phi}{r}\frac{\partial}{\partial\theta}\\\label{21}
&+&\frac{\csc\vartheta \cos\phi }{r} \frac{\partial}{\partial\phi}),
\\\label{22}
\textbf{Y}_{9}&=&r \cos\theta\frac{\partial}{\partial
t}+t(\cos\theta\frac{\partial}{\partial r}-\frac{\sin
\theta}{r}\frac{\partial}{\partial\theta}).
\end{eqnarray}

\section*{Appendix B}

For the orbital equation, the first-order approximate symmetries of
the Schwarzschild spacetime include
\begin{eqnarray}\label{61}
\textbf{Y}_{0}&=&u\frac{\partial}{\partial u },\quad
\textbf{Y}_{1}=\cos\phi\frac{\partial}{\partial u},\quad
\textbf{Y}_{2}=\sin\phi\frac{\partial}{\partial u}, \\\label{62}
\textbf{Y}_{3}&=&\frac{\partial}{\partial \phi },\quad
\textbf{Y}_{4}=\cos2\phi\frac{\partial}{\partial \phi}-u
\sin{2\phi}\frac{\partial}{\partial u},\\\label{63}
\textbf{Y}_{5}&=&\sin2\phi\frac{\partial}{\partial \phi}+u
\cos{2\phi}\frac{\partial}{{\partial}u},\\\label{64}
\textbf{Y}_{6}&=&u\cos\phi\frac{\partial}{\partial \phi}-u^{2}
\sin\phi\frac{\partial}{\partial u}, \\\label{65}
\textbf{Y}_{7}&=&u\sin\phi\frac{\partial}{\partial \phi}+u^{2}
\cos\phi\frac{\partial}{\partial u},
\end{eqnarray}
and two non-trivial stable approximate symmetry generators are
\begin{eqnarray}\label{66}
\textbf{Y}_{a1}&=&\sin\phi \frac{\partial}{\partial
u}+{\epsilon}(2\sin\phi\frac{\partial}{\partial \phi}+u
\cos\phi\frac{\partial}{\partial u}) ,\\\label{67}
\textbf{Y}_{a2}&=&\cos\phi \frac{\partial}{\partial
u}-{\epsilon}(2\cos\phi\frac{\partial}{\partial \phi}-u
\sin\phi\frac{\partial}{\partial u}).
\end{eqnarray}

\end{document}